\documentclass[pra,aps,twocolumn,superscriptaddress]{revtex4}
\pdfoutput=1
\input{Qcircuit}

\usepackage{amssymb,bm}
\usepackage{amsthm}
\usepackage{amsmath}
\usepackage{graphicx}

\usepackage{algorithm}
\floatname{algorithm}{Protocol}

\def\be{\begin{equation}}
\def\ee{\end{equation}}

\newcommand{\degree}[1]{\ensuremath{#1^\circ}}
\newcommand{\nuc}[2]{\ensuremath{{}^{#1}\textrm{#2}}}
\newcommand{\Bone}{\ensuremath{\textrm{B}_{1}}}
\newcommand{\Tone}{\ensuremath{\textrm{T}_{1}}}
\newcommand{\Ttwo}{\ensuremath{\textrm{T}_{2}}}

\theoremstyle{definition}

\theoremstyle{theorem}

\theoremstyle{theorem}

\theoremstyle{definition}

\usepackage{times}
\usepackage{color}

\begin{document}

\title{Quantum correlations which imply causation}
\author{Joseph Fitzsimons}
\email{joe.fitzsimons@quantumlah.org}
\address{Centre for Quantum Technologies, National University of Singapore, 3 Science Drive 2, 117543, Singapore}
\author{Jonathan Jones}
\address{Clarendon Laboratory, Department of Physics, University of Oxford, Parks Road, Oxford OX1 3PU, U.K.}
\author{Vlatko Vedral}
\address{Centre for Quantum Technologies, National University of Singapore, 3 Science Drive 2, 117543, Singapore}
\address{Clarendon Laboratory, Department of Physics, University of Oxford, Parks Road, Oxford OX1 3PU, U.K.}
\address{Department of Physics, National University of Singapore, 2 Science Drive 3, 117542, Singapore}

\date{\today}
\begin{abstract}
In ordinary, non-relativistic, quantum physics, time enters only as a parameter and not as an observable~\cite{busch2002time}: a state of a physical system is specified at a given time and then evolved according to the prescribed dynamics. While the state can, and usually does, extend across all space, it is only defined at one instant of time, in conflict with special relativity where space and time are treated on an equal footing.  Here we ask what would happen if we defined the notion of the quantum density matrix for multiple spatial and temporal measurements. We introduce the concept of a pseudo-density matrix which treats space and time indiscriminately. This matrix in general fails to be positive for timelike separated measurements, motivating us to define a measure of causality that discriminates between spacelike and timelike correlations. Important properties of this measure, such as monotonicity under local operations, are proved. Two qubit NMR experiments are presented that illustrate how a temporal pseudo-density matrix approaches a genuinely allowed density matrix as the amount of decoherence is increased between two consecutive measurements.
\end{abstract}
\maketitle

Ever since the pioneering work of Bell~\cite{bell1964einstein}, the study of quantum correlations has proved fertile ground for gaining insight into fundamental physics. Much of that progress has been focused on spatial correlations, in the form of entanglement and quantum discord~\cite{henderson2001classical,ollivier2001quantum,dakic2010necessary}, but a number of authors have extended this approach into the time domain. In particular Leggett and Garg showed that quantum systems exhibit a form of timelike correlation which cannot be accounted for by any macro-realistic theory~\cite{leggett1985quantum}. Similarly it has been shown that assumptions of realism and locality in time lead to a form of temporal Bell inequality, which again can be violated by quantum systems~\cite{brukner2004quantum}. Here we take a different approach: assuming quantum mechanics \textit{a priori}, and examining the correlations which can arise. Quantum states which violate the Leggett--Garg inequality necessarily exhibit the causal correlations we identify, and hence recent experimental demonstrations of violations of such inequalities may constitute a limited observation of such causal correlations~\cite{knee2012violation,dressel2011experimental,palacios2010experimental,goggin2011violation,waldherr2011violation}.

In quantum mechanics, each system in a multi-system quantum state is assigned a separate Hilbert space and these spaces are connected through the tensor product structure. The tensor product indicates that these systems are to be treated separately, though the joint state is also well defined at each instant of time. Here we explore extending this notion to different instances in time and assigning a Hilbert space to each different instant in time in much the same way as it is done in space. The resulting spatio-temporal state is then investigated. 

First we introduce the standard density matrix in quantum physics for qubits, although our ideas apply to subsystems of any dimensionality. We then show how to extend the concept of the spatial density matrix to different instances in time. The difference between spatial and temporal correlations is investigated through the introduction of the causality monotone, which is meant to capture the degree of ``temporalness'' in any quantum correlations. Finally, we present experiments using an NMR implementation that illustrate the basic properties of the pseudo-density matrices. Our result shows that the simple phrase ``correlation does not imply causation'', frequently heard among philosophers of science, is incorrect if taken at face value. Our proposal to treat spatial and temporal correlations within the same quantum formalism clearly still discriminates between the two, albeit imperfectly: when the pseudo-density matrix fails to be positive, this means that it necessarily contains a temporal element; the converse of this is not true, as the pseudo-density can be positive without implying spacelike separation.

The density matrix can be viewed as a probability distribution over pure states, with $\rho = \sum_i p_i \ket{\psi_i}\bra{\psi_i}$, where $p_i$ is the probability of the pure state $\ket{\psi_i}$ occurring. Given a density matrix $\rho$, the expectation value of a particular Pauli operator $P$ is $\langle P \rangle = \mbox{Tr}(P \rho)$. As the $n$ qubit Pauli operators along with the identity form a basis for the space of Hermitian operators, and any density matrix $\rho$ is necessarily Hermitian, it follows that any $\rho$ can be written as $\rho = a_o \mathbb{I} + \sum_i a_i P_i$, where $P_i$ is the $i$th Pauli operator on $n$ qubits, and $\{a_i\}_{i=0}^{n^2 - 1}$ are real numbers. Further, since Pauli operators are traceless, and all density matrices have unit trace, we have $a_0=1/2^n$ and the expectation value for $P_j$ is then given by
\begin{equation}
\langle P_j \rangle = \mbox{Tr}\left(P_j \left(\frac{\mathbb{I}}{2^n} + \sum_i a_i P_i\right)\right) = 2^n a_j.
\end{equation}
Thus we have an alternate formulation of the density matrix in terms of the expectation value of Pauli operators,
\begin{equation}
\rho = \frac{\langle \mathbb{I} \rangle}{2^n} \mathbb{I} +  \sum_{i=1}^{n^2 -1} \frac{\langle P_i \rangle}{2^n} P_i.
\end{equation}
As we are interested in discussing correlations, we can express each $n$ qubit Pauli operator as the product of single qubit operators, yielding
\begin{equation}
\rho = \frac{1}{2^n} \sum_{i_1 = 0}^{3} ... \sum_{i_n = 0}^{3} \left \langle \bigotimes_{j=1}^n \sigma_{i_j} \right \rangle \left(\bigotimes_{j=1}^n \sigma_{i_j} \right)
\end{equation}
where $\sigma_0 = \mathbb{I}$, $\sigma_1 = X$, $\sigma_2=Y$ and $\sigma_3 = Z$.

This above equation can be taken as definig a generalization of the density matrix. We consider a set of events $\{E_1 ... E_N\}$, where at each event $E_j$ a measurement of a single qubit Pauli operator $\sigma_{i_j} \in \{\sigma_0, ..., \sigma_3\}$ can be made. For a particular choice of Pauli operators $\{\sigma_{i_j}\}_{j=1}^n$, we take $\langle \{\sigma_{i_j}\}_{j=1}^n \rangle$ to be the expectation value of the product of the result of these measurements. Then we can define a pseudo-density matrix
\begin{equation}
R = \frac{1}{2^n} \sum_{i_1 = 0}^{3} ... \sum_{i_n = 0}^{3} \left \langle\{\sigma_{i_j}\}_{j=1}^n \right \rangle \bigotimes_{j=1}^n \sigma_{i_j}.
\end{equation}

If the measurement events $E_1 ... E_n$ are spacelike separated, then $R$ reduces to the standard $n$ qubit density matrix. However, as there is no notion of separate systems inherent in the definition of $R$, it allows us to describe correlations between measurement events which are not spacelike separated, for example encapsulating the possibility of multiple measurements made at different points in time on a single system. This is a generalization of the notion of a quantum state extended across spacetime, rather than the usual restriction to some fixed time. We note that Isham \cite{isham1995quantum} also considered introducing a direct product structure into temporal correlations, but within a completely different context.

This pseudo-density matrix inherits some properties of a standard density matrix. Firstly, it is Hermitian, since it is defined as a sum over Pauli operators, and secondly it has unit trace, since the expectation value $\langle \{I,...,I\}\rangle = 1$. Lastly, the expectation value for the product of any choice of local measurements $\{\sigma_{i_j}\}_{j=1}^n$ is given by
\begin{equation}
\left\langle\{\sigma_{i_j}\}_{j=1}^n\right\rangle = \mbox{Tr}\left( \left(\bigotimes_{j=1}^n \sigma_{i_j} \right)R\right).
\end{equation}
As with a standard density matrix, we can also trace over subsystems to produce a reduced pseudo-density matrix, defined only over the remaining events.

All density matrices are positive semi-definite matrices with unit trace, and any matrix satisfying these requirements can be interpreted as a density matrix. The main difference between $R$ and a standard density matrix, then, is that $R$ is not necessarily positive semi-definite. To see this, we consider the case of a single physical qubit with two separate measurement events. We take the qubit to be initially in the state $\ket{0}$ and assume that evolution between measurement events corresponds to the identity operator. In this case the expectation values are all zero, except for $\langle \{\mathbb{I},\mathbb{I}\}\rangle$, $\langle \{X,X\}\rangle$, $\langle \{Y,Y\}\rangle$, $\langle \{Z,Z\}\rangle$, $\langle \{Z,\mathbb{I}\}\rangle$, and $\langle \{\mathbb{I},Z\}\rangle$, which are all equal to one. From these expectation values, we obtain a pseudo density matrix
\begin{equation}
R = \left( \begin{array}{cccc}
1 & 0 & 0 & 0 \\
0 & 0 & \frac{1}{2} & 0\\
0 & \frac{1}{2} & 0 & 0 \\
0 & 0 & 0 & 0
\end{array} \right),
\end{equation}
which has eigenvalues $\{-\frac{1}{2}, 0, \frac{1}{2}, 1\}$. The existence of negative eigenvalues implies that $R$ is not positive-semi definite.

Any $R$ which is positive semi-definite can be interpreted as a regular density matrix, for which it is possible to duplicate the correlations present with spacelike separated quantum systems. However, when $R$ has negative eigenvalues, it cannot be interpreted as a regular density matrix, implying that the measurements cannot be spacelike separated, and hence there must exist a causal relation between events.

The causal relationship embodied in certain pseudo-density matrices has may similarities to another form of uniquely quantum correlation, entanglement, and we can define an analogous measure of causal correlations. In order for a function $f(R)$ to be considered a causality monotone we require the following criteria to hold:
\begin{enumerate}
\item $f(R) \geq 0$, with $f(R)=0$ if $R$ is completely positive, and $f(R_2) = 1$ for any $R_2$ obtained from two consecutive measurements on a single qubit closed system,
\item $f(R)$ is invariant under unitary operations,
\item $f(R)$ is non-increasing under local operations, and
\item $\sum_i p_i f(R_i) \geq f(\sum_i p_i R_i)$.
\end{enumerate}
These criteria correspond almost exactly to the criteria for an entanglement monotone \cite{wei2003maximal,vidal2000entanglement}, except that criterion three is somewhat weakened. An entanglement monotone is required not to increase on average under local operations and classical communication, however any processing based on classical communication would constitute a causal relationship and hence is excluded.

As we have shown, any pseudo-density matrix which embodies some form of causal relationship must have at least one negative eigenvalue. Since such matrices are Hermitian (and hence have real eigenvalues) and have unit trace, it follows that the trace norm is strictly greater than one. On the other hand, if all eigenvalues are positive, the trace norm is exactly one.

This leads us to define a measure based on the trace norm,  $f_{tr}(R) = ||R||_{tr} - 1$. As we have seen, $||R||_{tr} \geq 1$ for all valid pseudo-density matrices, and hence $f_{tr}(R) \geq 0$. Further, $f_{tr}(R)=0$ trivially for all positive semi-definite $R$, and from the previous example it is clear that  $f_{tr}(R_2)=1$ for at least one choice of $R_2$. Since the trace norm is unitarily invariant, the first and second criteria for $f_{tr}$ to be a causality monotone are satisfied. Similarly, by applying Stinespring dilation to represent local quantum operations as unitary operations on a larger Hilbert space, the third criterion follows directly since the trace norm is non-increasing under partial trace. The final criterion follows from the triangle inequality since $||\sum_i p_i R_i ||_{tr} - 1 \leq \sum_i p_i (||R_i||_{tr}-1)$ and hence $f_{tr}(\sum_i p_i R_i) \leq \sum_i p_i f_{tr}(R_i)$. Thus $f_{tr}$ is a causality monotone.

\begin{figure*}[!htb]
\centering
\includegraphics[width=\textwidth]{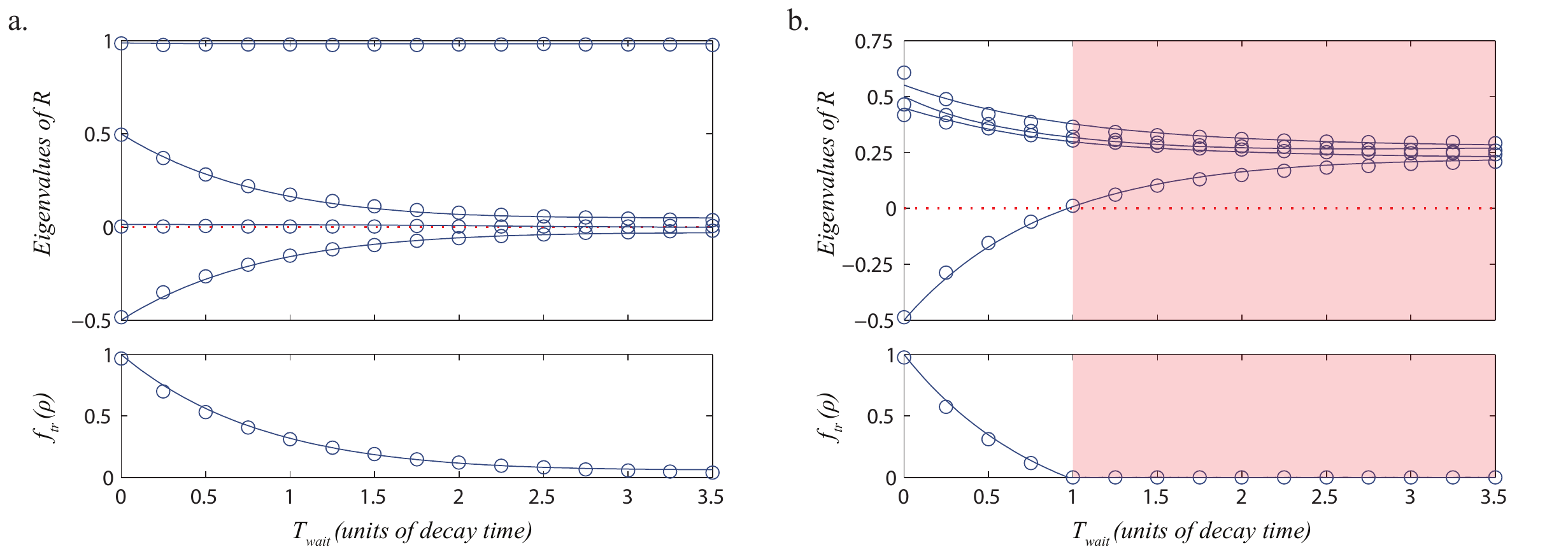}
\caption{\label{fig:results} Eigenvalues of $R$ and value of $f_\text{tr}$ as a function of $T_\text{wait}$. In (a) the system starts in a pseudo-pure state and undergoes dephasing noise, while in (b) the system starts in a mixed state and undergoes depolarising noise. The circles indicate data points obtained from experiment, while the solid lines indicate the best fit for the relevant theoretical models. These models each take 3 parameters to describe the initial state of the system and either 1 and 3 parameters, respectively, to parametrize the noise. The red region indicates the time period in which all resulting pseudo-density matrices are acausal.}
\end{figure*}

Naively it would appear that non-destructive single qubit measurements are necessary in order to perform the multi-event measurements required for tomography of a pseudo-density matrix. This would rule out the possibility of reconstructing a pseudo-density matrix in either NMR or quantum optics, two of the most established testbeds for quantum physics. Fortunately, however, it is possible to circumvent the limitations imposed by ensemble measurements by making use of an ancilla qubit to record the parity of the local Pauli measurement results.  Thus it is possible to recover their product by measuring a single spin, similar to the approach advocated in \cite{souza2011scattering}.

The simplest system for which $R$ can have negative eigenvalues contains two measurement events. These can be made either on the same qubit or seperate qubits. However in order to observe both causal and acausal correlations in the current generation of experiments we focus on measurements separated by a variable time on a single qubit,  as in the circuit below which accomplishes the measurement $\{\sigma_1,\sigma_2\}$.
\[
\centerline{
\Qcircuit @C=0.5em @R=0.5em {
&\lstick{\ket{0}}&\qw & \targ & \qw &\qw & \qw& \targ & \qw &\meter \\
&\lstick{\ket{0}}&\gate{U_{\sigma_1}}& \ctrl{-1} &\gate{U_{\sigma_1}^\dagger} & \gate{T_\text{wait}}  & \gate{U_{\sigma_2}}& \ctrl{-1} &\gate{U_{\sigma_2}^\dagger} & \qw \\
}}
\]
Here $U_{\sigma_{1(2)}}$ is the unitary operation mapping the $\pm 1$ eigenstate of the Pauli operator $\sigma_{1(2)}$ onto the $\pm 1$ eigenstate of $Z$. Between measurements we allow a period of free evolution, during which the primary qubit undergoes decoherence, and we calculate a pseudo-density matrix $R_{T_\textrm{wait}}$ for a range of waiting times.

NMR experiments were performed on a Varian Unity Inova spectrometer with a nominal \nuc{1}{H} frequency of 600\,MHz using a HF\{CP\} probe with pulsed field gradients.  The NMR sample comprised \nuc{13}{C}-labelled sodium formate dissolved in $\textrm{D}_2\textrm{O}$ at \degree{20}C, providing a heteronuclear two-spin system.  The \nuc{1}{H} spin was used as the primary qubit and the \nuc{13}{C} spin as the ancilla. Both spins were placed on resonance, so that the Hamiltonian took the form of a spin--spin $ZZ$ coupling of 194.7\,Hz, and the \Bone\ field strengths were adjusted to give nutation rates of 12.5\,kHz.  The measured relaxation times were \Tone=7.8\,s and \Ttwo=3.2\,s for \nuc{1}{H} and \Tone=16.3\,s and \Ttwo=6.7\,s for \nuc{13}{C}.  An inter-scan delay of 60\,s ensured that the spin system began each experiment close to its thermal state.

Quantum logic gates were implemented using standard approaches \cite{Jones2001a,Jones2011}.  Single qubit rotations in the $XY$-plane were implemented using BB1 composite rotations \cite{Wimperis1994,Cummins2003}, while $Z$-rotations were implemented as frame rotations \cite{Knill2000} which were propagated through the pulse sequence \cite{Bowdrey2005} to points where they could be dropped.  Pseudo-pure two-qubit states were prepared using the method of Kawamura \textit{et al.} \cite{Kawamura2010}; for pseudo-pure single qubit states the thermal state was used directly.  NMR spectra were processed using home written software and the intensity of the \nuc{13}{C} doublet determined by combining separate integrals for the two components; all integrals were normalised using a reference spectrum.

Instead of using natural decoherence during $T_\text{wait}$, controllable dephasing of the primary qubit was implemented using the diffusive suppression of pulse field gradient spin echoes \cite{stejskal1965} as described by Cory \textit{et al.} \cite{Cory1998}.  This can be converted to controlled depolarization by using single qubit rotations to apply the dephasing around the $X$, $Y$ and $Z$-axes in turn.  This process also dephases the ancilla qubit, but leaves its $Z$-component unaffected as the ancilla does not experience the single qubit gates.

Figure \ref{fig:results} shows the results of our NMR experiments, plotting the eigenvalues of the pseudo-density matrices as a function of time, along with the corresponding $f_{tr}$, in each of two settings. Figure \ref{fig:results}A shows the results of purely dephasing noise acting on an initial state pseudo-pure state $\ket{0}$.  Here the pseudo-density matrix starts with a single negative eigenvalue, which tends towards zero from below as the waiting time is increased. The pseudo-density matrix never becomes positive semi-definite (and hence acausal) because the decoherence brings it towards a matrix which is rank deficient, and so the minimum eigenvalue approaches, but never quite reaches, zero.

In order to observe a sharp transition between causal and acausal pseudo-density matrices it is necessary both to start with a mixed initial state, and to allow depolarizing decoherence, which is the case considered in Figure \ref{fig:results}B. Now we observe a transition between causal and acausal pseudo-density matrices as the minimum eigenvalue crosses the zero threshold, a phenomenon reminiscent of entanglement sudden death~\cite{yu2009sudden}.

\textit{Acknowledgements -- }We thank John Baez, Andreas Doering and Pieter Kok for helpful discussions. JF and VV acknowledge support from the National Research Foundation and the Ministry of Education, Singapore. VV also thanks the James Martin School (UK), Leverhulme Trust (UK) and the Templeton Foundation (USA).

\bibliographystyle{apsrev}
\bibliography{caus}

\end{document}